\documentclass[twocolumn]{aastex631}
\usepackage{float}
\usepackage{amsmath}

\shorttitle{RMHD PINNs}
\shortauthors{Cheung et al.}

\begin{document}

\title{Reconstructing Relativistic Magnetohydrodynamics with Physics-Informed Neural Networks}

\affiliation{Harvard College, Cambridge, MA 02138, USA}
\affiliation{Harvard John A. Paulson School of Engineering and Applied Sciences,
150 Western Avenue, Allston, MA 02134, USA}
\affiliation{Center for Astrophysics | Harvard \& Smithsonian,
60 Garden Street, Cambridge, MA 02138, USA}
\affiliation{Institute for Advanced Study, 1 Einstein Drive,
Princeton, NJ 08540, USA}
\affiliation{OpenAI, 1455 3rd Street, San Francisco, CA 94158, USA}

\author{Corwin Cheung}
\affiliation{Harvard College, Cambridge, MA 02138, USA}
\affiliation{OpenAI, 1455 3rd Street, San Francisco, CA 94158, USA}

\author{Marcos Johnson-Noya}
\affiliation{Harvard College, Cambridge, MA 02138, USA}
\affiliation{Harvard John A. Paulson School of Engineering and Applied Sciences, 150 Western Avenue, Allston, MA 02134, USA}

\author{Michael Xiang}
\affiliation{Harvard College, Cambridge, MA 02138, USA}
\affiliation{Harvard John A. Paulson School of Engineering and Applied Sciences, 150 Western Avenue, Allston, MA 02134, USA}

\author{Dominic Chang}
\affiliation{Center for Astrophysics | Harvard \& Smithsonian, 60 Garden Street, Cambridge, MA 02138, USA}

\author{Alfredo Guevara}
\affiliation{Institute for Advanced Study, 1 Einstein Drive, Princeton, NJ 08540, USA}

\begin{abstract}
We construct the first physics-informed neural-network (PINN) surrogates for relativistic magnetohydrodynamics (RMHD) using a hybrid PDE and data-driven workflow. Instead of training for the conservative form of the equations, we work with Jacobians or PDE characteristics directly in terms of primitive variables. We further add to the trainable system the divergence-free condition, without the need of cleaning modes. Using a novel MUON optimizer implementation, we show that a baseline PINN trained on early-time snapshots can extrapolate RMHD dynamics in one and two spatial dimensions, and that posterior residual-guided networks can systematically reduce PDE violations. 
\end{abstract}

\section{Introduction}\label{sec:intro}

Electromagnetic observations of various astrophysical phenomena are described by emission from relativistic plasmas.
Accurate descriptions of these systems are necessary to answer longstanding questions in astronomy, such as, how a galaxy's structure is influenced by its central supermassive black hole \cite{hyerin,Joana}, how supermassive black holes grow and evolve \cite{Ricarte_2023}, and the mechanisms behind pulsar and quasar emission  \cite{McKinney_Pulsar,Ighina_quasar}. 
In many of these systems, the bulk flow of the plasma can reach relativistic speeds, and is well described as a conducting fluid whose motion is constrained by the presence of a magnetic field.
The framework that describes these assumptions is that of Relativistic magnetohydrodynamics (RMHD).

RMHD reduces to the case of ideal RMHD when plasmas are assumed to have negligible resistivity. 
The governing equations for ideal RMHD are a set of eight hyperbolic first order partial differential equations (PDEs) with an elliptic constraint.
The structure of these equations permits the application of various conservative schemes for generating solutions \citep{Toro2009-ty}, which are necessary to accurately capture shocks that arise generically during evolution.
These schemes, however, all feature characteristic deficiencies which they must overcome.
Firstly, the requirement of consistency with relativity requires the definition of suitable inversion algorithms between primitive and conserved quantities \cite{Noble:2006pvs}.
There is no unique way to define this inversion, with many approaches incapable of guaranteeing convergence.
Secondly, the constraint equations impose an additional requirement that must be handled independently from how the set of hyperbolic PDEs are solved.
Lastly, the conditions that determine numerical convergence are decided at the beginning of a simulation run, and cannot be changed to improve convergence after a solution has been generated.

Several conservative schemes for RMHD have been proposed with different levels of accuracy, \citep[see e.g.][]{Porth:2017BHAC,Gammie:2003HARM,Sadowski:2013KORAL}. 
Such schemes are plagued by high computational expense that imposes practical limitations on reaching numerical convergence.
For example, many schemes require the implementation of artificial numerical floors to control for low-plasma density regions \citep[see for example][]{chael_force_free} and are inadequate for handling deviations from the case of ideal RMHD due the stiffness imposed by additional terms \citep[][]{Ripperda_2019}.
An inherent flaw to such schemes is that their accuracy is determined from the initial set-up, which prevents future corrections once approximate solutions are generated.

Machine learning methods could deliver the potential means to generate accurate solutions to these equations more efficiently. 
The current advent of physics-informed neural-networks (PINNs), powered by advances in neural network architecture and optimization techniques propose an appealing alternative to existing conservative schemes. 
PINNs have been successfully applied to non-relativistic MHD problems \citep{NASA_PINN,Dimitropoulos:2025pulsarML}, RHD problems \citep{FerrerSanchez2024GAPINN}, as well as a wide class of non-linear PDE systems \citep[][]{RAISSI2019686, self_similar_euler}. 

Motivated by this, in this work we explore the implementation of a RMHD solver through PINNs.
Our method addresses some of the downfalls of traditional conservative schemes. Using a state-of-the-art optimizer scheme (MUON) our method is able to achieve convergence to high precision through continuous training, and to solve for primitive quantities directly by bypassing the inversion scheme necessary for traditional integrators to work. 
Through hyperparameter search we find a suitable architecture to be trained in diverse RMHD tests in one and two dimensions. 
Furthermore, in order to guarantee convergence of the network we implement a data driven training during early times, and a PDE/residual training at late times. 

\section{From Conservative Scheme to Characteristics}

The equations of ideal-RMHD follow from stress-energy conservation, mass-current conservation, and Maxwell's equations under the ideal-MHD condition of no electric field in the frame of the fluid. 
They are a set of $8$ hyperbolic PDEs with an additional parabolic constraint equation.
They are often written in the form,
\begin{align}
  \partial_t \mathcal{U}^a(\mathcal P) + \partial_i\mathcal{J}^{a\,i}(\mathcal P) &= 0,   \label{eq:consd} \\
  \partial_i B^i
    &=0.\label{eq:dv0}
\end{align}
The first term corresponds to the conserved variables,
\begin{equation}
    \mathcal U=(D,\varepsilon,m_i,B_i),
\end{equation} 
which are the relativistic mass density in the laboratory frame ($D$), the total energy density ($\varepsilon$), momentum density ($m_i$),  and the laboratory frame magnetic field, ($B^i$). 
The $\{i,j\dots\}$ indexes range over $\{x,y,z\}$ spatial directions. 
We use the first half of the Latin alphabet $a,b=1,\ldots ,8$ to index entries associated with conservative variables and primitives.

The conserved quantities are explicitly known functions of the primitives,
\begin{equation}
  \mathcal{P} = (\rho, p, v^i,  B^i).
\end{equation}
which stand for laboratory frame density, pressure, velocity and magnetic field respectively. 
The explicit form of the functions $\mathcal U(\mathcal P)$ and $\mathcal J(\mathcal P)$ is well known for ideal RMHD and General Relativistic magnetohydrodynamics (GRMHD). 
We provide the former in a conventional notation in Appendix \ref{app:jacobians}. 
We assume an equation of state parametrized by an adiabatic index $\Gamma$,
\begin{equation}
    p = (\Gamma-1) u
\end{equation}
where $u$ is the internal energy. Throughout this work different values of $\Gamma$ will be considered. 

Conservative schemes for solving GRMHD equations are plagued by inescapable difficulties.
Firstly, the function $\mathcal P(\mathcal U) = \mathcal U^{-1}(\mathcal P)$ is not bijective, and requires an iterative  approach for inversion in conservative schemes.
These procedures usually require the approximation of roots of high degree polynomials at every time step. 
Later, the function is used to obtain the currents $\mathcal J^i$ which are used to approximate their flux exchange between cells of finite volume. 
This step is typically approximated by averaging quantities within volumes, and solving the transfer of information between cells as a Riemann problem.
Another well known limitation of the system is the preservation of the divergence-free condition $\partial_iB^i =0$ which is crucial for proper time evolution. \footnote{See, e.g., \cite{Owkes2024FirstCFDSolver,Toro2009} for an introductory outline of numerical CFD methods.}

The computational cost of the above procedure calls for an alternative approach which bypasses the inversion scheme $\mathcal U\to \mathcal P(\mathcal U)$ or the strict hyperbolic requirement. 
Many of the difficulties associated with conservative schemes are bypassed with non-conservative schemes that avoid the evolution of conserved quantities.
However, these schemes can often be unstable or lead to spurious solutions which are harder to interpret.
The RMHD-NN code exploits the linear structure of the PDE system by working directly with its characteristics for the primitive variables, without reference to conserved quantities or fluxes. 
Indeed, we start by observing that \eqref{eq:consd} is equivalent to the following system
\begin{equation}\label{eq:lnsq}
  \mathcal M_{ab}(\mathcal P)\,\partial_t \mathcal P^b + \mathcal A_{ab}^{\,i}(\mathcal P)\partial_i\mathcal P^b = 0,
\end{equation}
where
\begin{equation}
  \mathcal M_{ab} = \frac{\partial \mathcal U^a}{\partial\mathcal P^b}, \qquad
   \mathcal A_{ab}^{\,i} = \frac{\partial \mathcal J^{a\,i}}{\partial \mathcal P^b}.
\end{equation}
where $\{a,b\}$ indexes through the entries associated with each conserved quantity/primitive.
The explicit form of these Jacobians is quoted in the Appendix. Schematically, $\mathcal M^{-1}\mathcal A=\frac{\partial\mathcal J}{\partial\mathcal U}$ is the usual characteristic matrix. 
Positive eigenvalues reflect the hyperbolicity of the system, but can be spoiled due to failure of the constraint \eqref{eq:dv0}. 
As we will not be concerned with time evolution in our method we find this is not needed. Instead, we propose to simply augment the system \eqref{eq:lnsq} to \textit{9 equations}, the last one being purely spatial, the divergence-free constraint, improving convergence.

\begin{figure*}[t]
\centering
\makebox[\textwidth][c]{%
  \includegraphics[width=\textwidth]{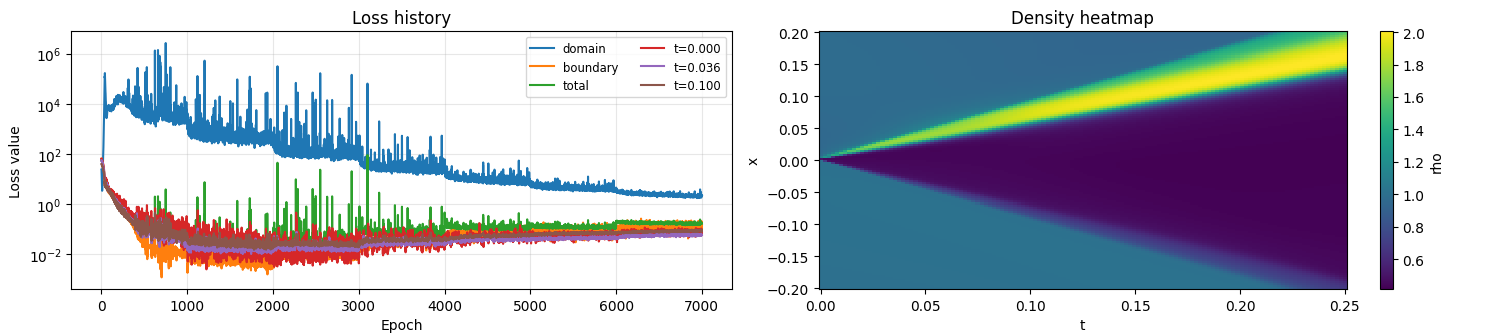}%
}
    \caption{
    \textit{Representative 1D training process for 1D RMHD shocktube}. 
    The left panel  shows the training convergence of the total loss function (green) and its weighted constituents --- contributions from the boundary condition (orange) and data loss at times $t{=}0.0$ (red), $t{=}0.036$ (purple) and $t{=}0.10$ (brown). 
    The unweighted domain loss (blue) is also shown. 
    The right panel shows the mass density $\rho$ over the domain as predicted by the model.}
    \label{fig:train1d}
\end{figure*}

In general, the characteristics of MHD systems are physically relevant for the following reason (see, e.g., \cite{Anton2008PhD,AntonEtAl2010,SchoepeHilditchBugner2018,HilditchSchoepe2019,Friedrichs1974,Anile1989}): Linearizing the previous system around a homogeneous background, and keeping only derivative terms, yields
\begin{equation}
\mathcal M_{ab}(\mathcal P)\,\partial_t \delta \mathcal P^{b}
\;+\;
\mathcal A_{ab}^{i}(\mathcal P)\,\partial_i \delta \mathcal P^{b}
\;\approx\; 0
\end{equation}
which admits a wave solution, $\delta \mathcal P \propto \exp(i\omega t + i  k^i x_i)$, whose characteristic speeds are the eigenvalues of \(\mathcal M^{-1}\mathcal A_i k^i\). The hyperbolic nature of the system is reflected in a ``7-wave fan'' of characteristics propagating at different rates. For instance, the Alfv\'en family provides an important diagnostic: the Alfv\'en speed (not to be confused with fluid speed)
\begin{equation}\label{eq:alfven}
  v_A^2 = \frac{b^2}{b^2 + \rho h}
\end{equation}
depends on the magnetic-field energy density \(b^2\) in the fluid frame
\begin{equation}
    b^2 \;=\; \frac{B_i B^i}{\gamma^2} + (v_i\,B^i)^2
\end{equation}
where $\gamma$ is the Lorentz factor and \(\rho h = \rho + \Gamma u\) is the enthalpy, as detailed in Appendix \ref{app:jacobians}. Correctly reproducing these characteristics is a consistency check for the RMHD implementation. Note that extra waves can also emerge in some circumstances due to artificially enforcing the constraint \eqref{eq:dv0}, see \cite{Dedner:2002glm}, but in our setup it will not be necessary.

\section{Physics-Informed Neural Networks for RMHD}\label{sec:pinn}
Here, we detail the network architecture and training schedule used for our experiments.

\subsection{Architecture}

We represent the primitive variables by a neural surrogate,
\begin{equation}
  (x^\mu)\longmapsto\; \mathcal P(x^\mu) =
  (\rho(x^\mu), p(x^\mu), v^i(x^\mu), B^i(x^\mu)).
\end{equation}
where $x^\mu=(t,x^i)$. In practice this is a fully connected multilayer perceptron (MLP) with different widths and depths for 1D and 2D tests.
A final exponential layer is also included to normalize $p$ and $u$ to be positive and $v^2<1$.
In the 1D case we use about 24 layers with width 64; in 2D we use about 64 layers with width 128. The nonlinearities are trainable hyperbolic tangents \cite{JagtapKawaguchi}.  For general background on variational PINNs see, e.g., \cite{Kharazmi:2020vpinn}.

The total loss is a weighted sum which reads, schematically,
\begin{equation}
  L_{\rm tot} = w_1 L_{\rm domain} + w_2 L_{\rm data} + w_3 L_{\rm bdy},
\end{equation}
where:
\begin{itemize}
  \item \(L_{\rm domain}\), namely \textbf{domain PDE loss}, measures the RMHD Jacobian residual, evaluated at random spacetime collocation points. The residual is built directly from the L2 norm of \eqref{eq:lnsq} at the points (MSE). Note that $\partial_i \mathcal P$ includes $\partial_i B_j$, thus we can augment this equation by the divergence-free condition \eqref{eq:dv0} without extra compute. This will increase the accuracy of the solver.
  \item \(L_{\rm data}\) is the L2 norm fitting early-time simulation snapshots supplied in the \texttt{data1d} and \texttt{data2d} folders. This includes the initial slice at $t=0.0$ which would be the standard data in the initial value problem.
  \item \(L_{\rm bdy}\) enforces open boundary conditions by penalizing deviations of \(\mathcal P(x,t)\) from constant-in-time boundary values. In practice this means to preserve the boundary conditions that were imposed at $t=0.0$.
\end{itemize}

We find that supplying the early time data, and not just the $t=0.0$ snapshot, helps drive the early training towards the correct minimum of the PDE and not just a local one. As has been observed many times with PINNs, not enforcing initial conditions strongly enough leads to essentially trivial solutions $\mathcal P^a \approx 0$ which only minimize the domain loss.

\begin{figure*}[t]
  \centering
  \includegraphics[width=\textwidth]{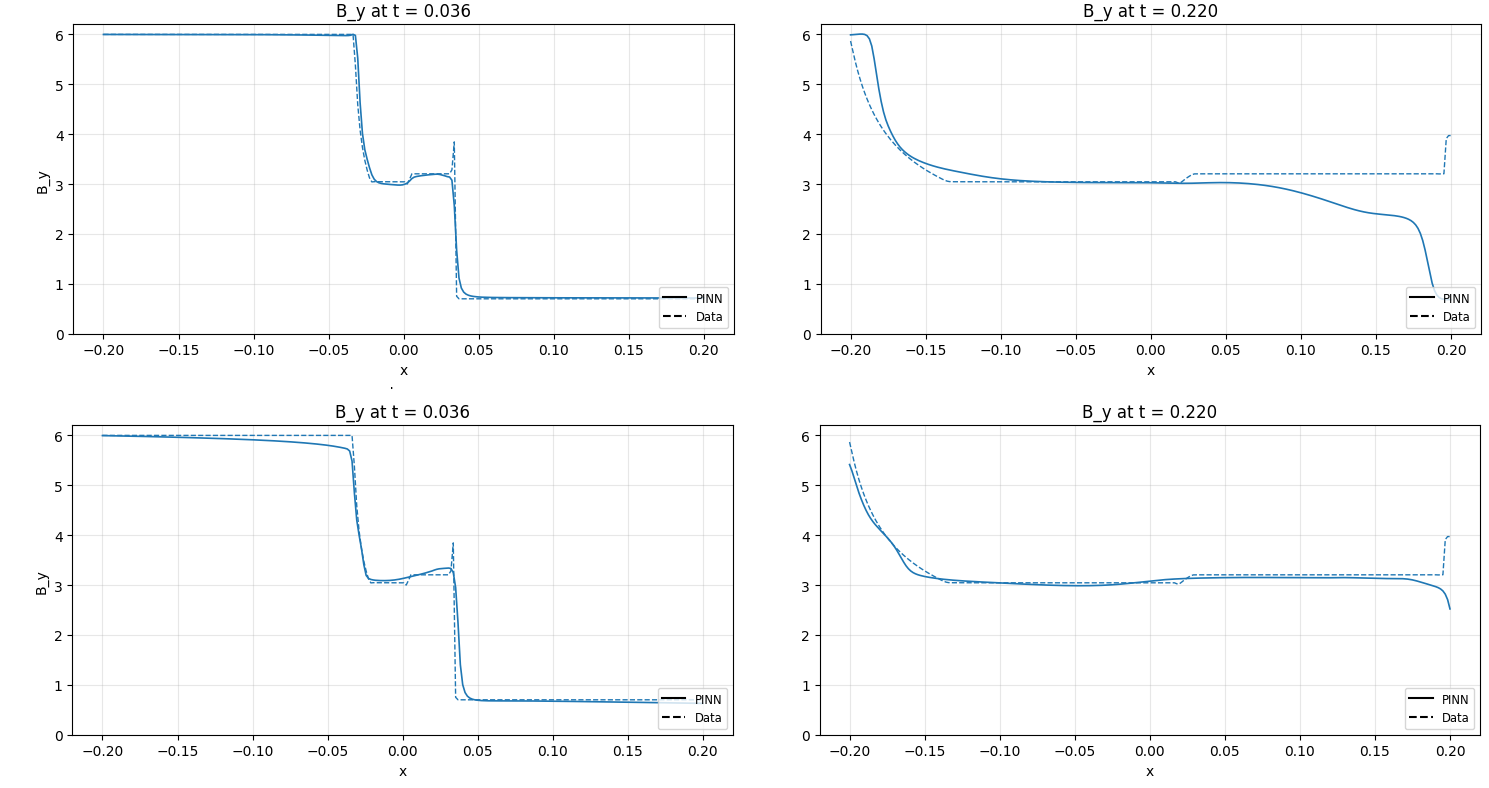}
  \caption{\textit{A tradeoff between fitting and extrapolating}.
  The two upper panels shows the network prediction early into training (${\sim} 1000$ epochs).
  We see good fitting of data conditions at $t{=}0.036$ (upper-left) and poor extrapolation to $t{=}0.22$ (upper-right). 
  The bottom two panels shows how the networks behaviour changes as the training progresses (${\sim} 7000$ epochs). 
  Now, the quality of the network's fit to the data conditions slightly loosen, but  the networks extrapolation quality increases. }
  \label{fig:tradeoff}
\end{figure*}

Once the baseline PINN fits the initial data accurately, it extrapolates the shock evolution to later times where no data are given and the training signal is entirely from the domain PDE residual. 
We refer to this process as \textit{tradeoff between fitting and extrapolation}, in direct analogy with the well known bias-variance tradeoff, which will be exemplified explicitly in the following numerical experiments.

\subsection{Training}
Previous studies have suggested that training benefits from learning scheduling which initially bias boundary data and initial data \citep{wang_pinn_guide}.
We implement such a training regimen, by dynamically increasing the PDE weight $w_1$ every ${\sim}10^3$ epochs. 
We also increase the number of domain collocation points and decrease the learning rate to control for stochastic behavior in the loss function.
To prevernt over-fitting, we randomly re-sample the domain, boundary and data slices on each epoch.
Finally, we use the MUON optimizer \citep{jordan2024muon}, which we find during our tests to produce the fastest training in comparison to LBFGS and ADAM \citep{LBFGS, ADAM}. 

A representative such training process is illustrated in Figure \ref{fig:train1d}.
We note that our training schedule artificially causes the total loss to increase even though the network actually learns the solution over the course of training (see \autoref{fig:tradeoff} for an illustrative example).
Our main focus is the loss of the unweighted PDE loss function which enables the network to have better extrapolation. 

\section{Numerical Experiments}\label{sec:results}

Here we will examine the convergence of the code for various classic tests of RMHD. Various 1D and 2D problems exhibit shockwaves characteristic of the dispersion relations of MHD. The tests we run are surveyed in \cite{Nagataki:2009grmhd}.
\subsection{1D shocktube}

Our preliminary test is the 1D shock-tube test. It features a pressure gradient propagating in both directions at different velocities. Early-time data are supplied at
\begin{equation}
  t = 0.0,\; 0.036,\; 0.1,
\end{equation}
and used only in \(L_{\rm data}\). We will use the ideal gas EOS with
\begin{equation}
    \Gamma= 5/3
\end{equation} 
Further initial data is described by Table \ref{tab:1dshock_filled}.

\begin{table}[H]
\centering
\caption{Initial conditions (primitive variables) for the 1D RMHD shock-tube at \(t=0\).}
\label{tab:1dshock_filled}
\begin{tabular}{lcccccccc}
\hline
Region & $\rho$ & $v^x$ & $v^y$ & $v^z$ & $B^y$ & $B^z$ & $p$ & $B^x$ \\
\hline
Left  & $1.0$ & $0$ & $0$ & $0$ & $6.0$ & $6.0$ & $30$ & $5.0$ \\
Right & $1.0$ & $0$ & $0$ & $0$ & $0.7$ & $0.7$ & $1.0$ & $5.0$ \\
\hline
\end{tabular}
\vskip 4pt
\end{table}

The adaptation of the code for a 1D problem, rather than a 2D problem is straightforward, we simply proceed to mask the $x$-components of the velocity and magnetic fields. We set the constant
\begin{equation}
    B_x=5.0
\end{equation}
In the 1D case, this satisfies the divergence constraint \eqref{eq:dv0} trivially.

Training the 1D RMHD problem provides a good illustration of the tradeoff between fitting and extrapolation. Once the baseline initial data is fitted, gradually tunning the weights loosens the fit but extrapolates the shock evolution to later times $\sim 0.25$ (where no data are given) and the training signal is entirely from the PDE residual. The model then provides a genuine a prediction that can be contrasted with non-training data. This is depicted in Figure \ref{fig:tradeoff}.

On the other hand, figure~\ref{fig:train1d} shows a typical training history and a snapshot of the $B_y$ profile. Based on the $x-t$ evolution plot, we find that the model is able to reproduce 6 of the RMHD characteristics. Using the results at $t=0.25$ and the relation \eqref{eq:alfven}, we can estimate the propagation of the Alfv\'en shock to be
\begin{equation}
   v_A \approx 0.82
\end{equation}
in good agreement with the outer shockwave estimated from the $x-t$ plot in \autoref{fig:train1d}.

Next we analyze the outcome of two-dimensional tests.

\subsection{2D cylindrical explosion}

In two spatial dimensions we consider a cylindrical explosion in a magnetized background. This is a signature RMHD test which in our case will assess the model capacity to reproduce a cylindrically symmetric solution (such symmetry is not imposed by the network). An initially over-pressurized circular region expands into a magnetized background, generating an outward-moving shock. The initial setup is described in the table below.

\begin{table}[h!]
\centering
\caption{Initial Conditions for 2D RMHD Cylindrical Blast Explosion}
\label{tab:2dblast}
\begin{tabular}{cccccc}
\hline
Region & $\rho$ & $p$ & $v^x$ & $v^y$ & $B^x$ \\
\hline
Background 
& $1.0$ 
& $0.01$ 
& $0$ 
& $0$ 
& $4.0$ \\
Blast (inside $r \le 0.08$) 
& $1.0$ 
& $10^{3}$ 
& $0$ 
& $0$ 
& $4.0$ \\
\hline
\end{tabular}

\begin{flushleft}
\footnotesize
Note.—Cylindrical blast centered at $(0.5,0.5)$ on a $[0,1]\times[0,1]$. Initial data is simulated on a grid with $250\times250$ points.  
The equation of state uses $\Gamma = 4/3$.  
Final time is $t = 0.4$.
\end{flushleft}
\end{table}

Compared to the 1D test, the network uses a deeper (64 layers) and wider (128 nodes) MLP with trainable \(\tanh\) activations. Training follows the same staged schedule: early-time data fitting, followed by an increasing emphasis on the PDE residual.

\begin{figure}[t]
\centering
\makebox[\linewidth][c]{%
  \includegraphics[width=1.1\linewidth,height=3cm]{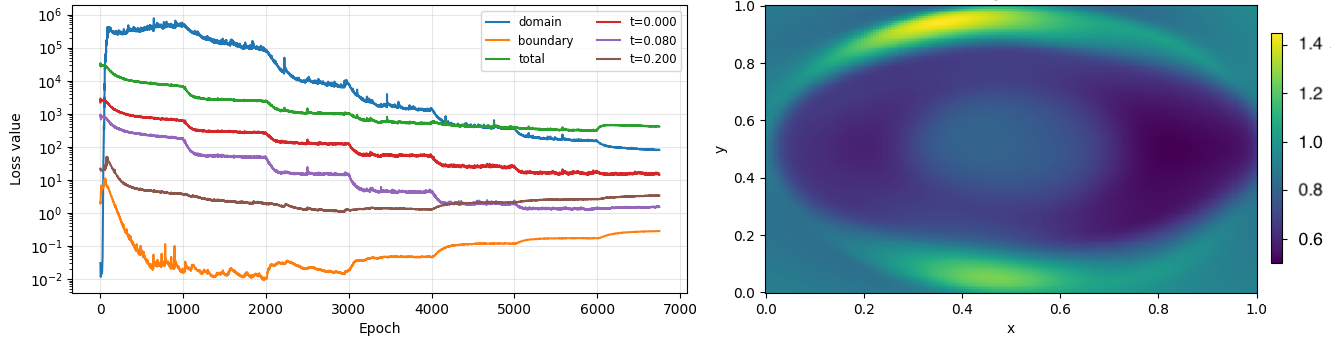}
}
  \caption{\textit{2D cylindrical test}. The left panel shows the evolution of the loss functions during training.
  Shown are the total loss function (green),  the weighted boundary loss (orange), the weighted data losses at times $t{=}0.0$ (red), $t{=}0.080$ (purple) and $t{=}0.20$ (brown), and the unweighted PDE loss (blue).
  The right panel is a heatmap of our PINN's prediction of the density at $t=0.40$. We refer to the RMHD-NN repository for more details.}
  \label{fig:train2d}
\end{figure}

\begin{figure}[t]
  \centering
  \includegraphics[width=\linewidth]{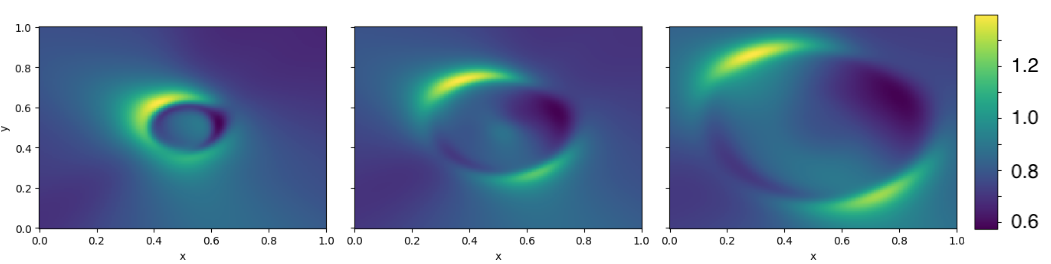}
  \caption{\textit{Density levels for 2D cylindrical explosion test}.  Model snapshots at $\textbf{t{=}0.0}$ (left), $\textbf{t{=}0.2}$ (middle) and $\textbf{t{=}0.4}$ (right).}
  \label{fig:cexp2d}
\end{figure}

Figure~\ref{fig:train2d} shows an example of the 2D loss evolution during training. Figure~\ref{fig:cexp2d} shows a representative snapshot of the cylindrical explosion.  

\begin{figure}[t]
  \centering
\includegraphics[width=1.0\linewidth , height = 0.5\linewidth ]{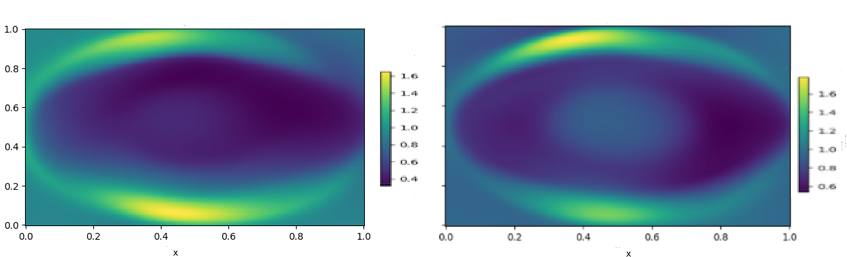}
\caption{Comparison before (left) and after (right) imposing $\partial_i B^i =0 $ constraint. We find that the constraint not only increases convergence but helps resolve the internal shock.}
  \label{fig:divfree_compare}
\end{figure}

\textit{Divergence free condition.} Crucially, we find that imposing the divergence free constraint \eqref{eq:dv0} as an extra PDE term significantly improves the resolution of this test, see Figure \ref{fig:divfree_compare}.

\subsection{2D shock tube}

As a final example we consider the 2D shock tube problem in relativistic hydrodynamics (RHD). This setup shows versatility of the 2D code to adapt to hydrodynamic problems in the absence of magnetism, $B_i=0$. We use the same parameters and network as in the cylinder explosion, but in this case the domain is partitioned into four quadrants with sharp discontinuities in between.

The expectation is that each interface generates a wave fan, but because interfaces meet at right angles, the resulting waves intersect and refract.

Note that $\Gamma$ in the equation of state is set to $5/3$. Final time is again set to $0.4$, with training data at $t=0, t=0.08$ and $t=0.2$. The details of the initial conditions per quadrant are given in Table \ref{fig:2dsckt}.
\begin{table}[h!]
\centering
\caption{Initial Conditions for 2D Shock Tube Problem}
\label{tab:2dshock}
\begin{tabular}{ccccccc}
\hline
Region & $x$ & $y$ & $\rho$ & $p$ & $v^x$ & $v^y$ \\
\hline
A & $0 \le x \le 0.5$ & $0.5 \le y \le 1$   & $0.1$   & $1$  & $0.99$ & $0$ \\
B & $0.5 \le x \le 1$ & $0.5 \le y \le 1$   & $0.1$   & $0.01$ & $0$    & $0$ \\
C & $0 \le x \le 0.5$ & $0 \le y \le 0.5$   & $0.5$ & $1$  & $0$    & $0$ \\
D & $0.5 \le x \le 1$ & $0 \le y \le 0.5$   & $0.1$ & $1$    & $0$    & $0.99$ \\
\hline
\end{tabular}
\label{fig:2dsckt}
\end{table}

As expected in the absence of electromagnetism, we find the network training converges extremely well in this case, reaching accuracies of $10^{-1}$ during a 90 minute training on a single GPU, see Figure \ref{fig:cexp2d}.

\begin{figure}[t]
  \centering
  \includegraphics[width=\linewidth]{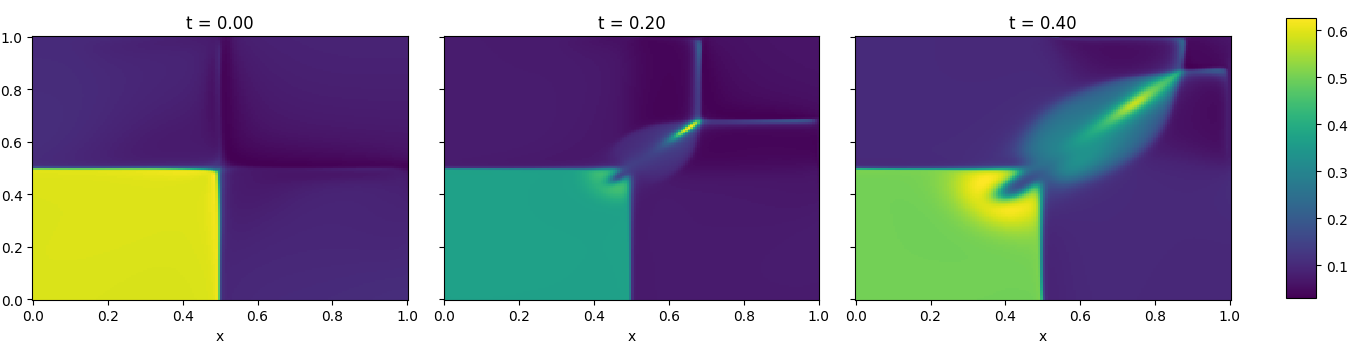}
  \caption{2D shocktube tube (pressure) reconstructed from the same architecture as the cylindrical test. The evolution shows a rarefaction oblique fan where the pressure decompresses. Model snapshots at $t{=}0.0$ (left), $t{=}0.2$ (middle) and $t{=}0.4$ (right).}
  \label{fig:shock2d}
\end{figure}

\begin{figure}[t]
  \centering
  \includegraphics[width=\linewidth]{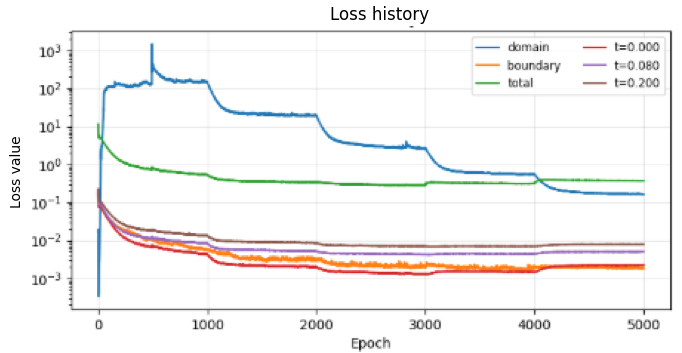}
  \caption{Training convergence of the 2D shocktube (note total loss is not normalized).}
  \label{fig:shock2d_train}
\end{figure}

\section{Residual-guided Correction Networks}\label{sec:residual}

Residual networks have recently emerged as a surrogate method to increase accuracy \cite{Wang:2025unsing}. They are based on a linearized version of the PDE, which in our case appears natural since it corresponds to well-known characteristics of MHD systems \cite{Teukolsky:2025charac}.

\begin{figure}[t]
  \centering
  \includegraphics[width=\linewidth]{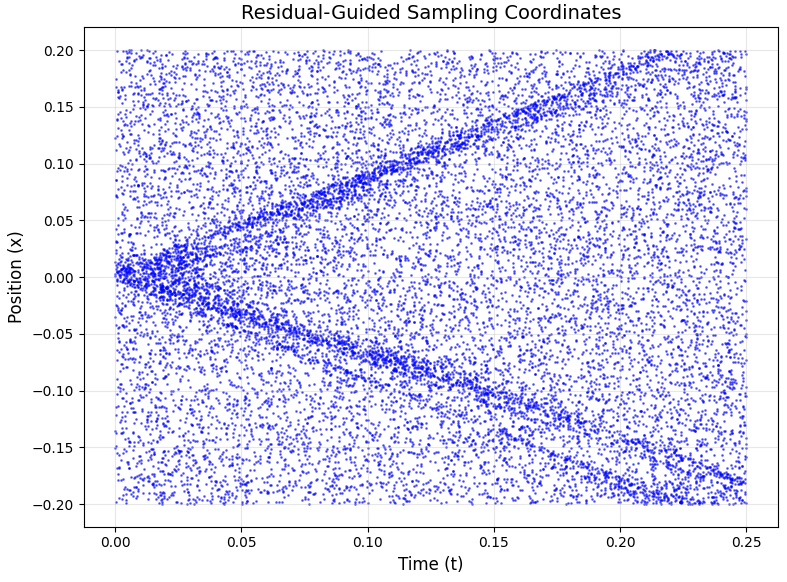}
  \caption{Monte-Carlo sampling for the 1D shock-tube test. 
  Points are concentrated where the Jacobian residual is largest.}
  \label{fig:residualsample}
\end{figure}

After the baseline model has converged, we evaluate the RMHD Jacobian residual on a new set of collocation points and construct a sampling density biased toward regions of large residual. For instance, in 1D we evaluate/store the residual
\begin{equation}
\mathcal{R}^{a}(\mathcal{P})
=
\mathcal{M}^{a}{}_{b}(\mathcal{P})\,\partial_t \mathcal{P}^{b}
+
A^{a}_{b\,x}(\mathcal{P})\,\partial_x \mathcal{P}^{b}
\end{equation}
and construct a training set of points through Monte Carlo sampling, see Figure \ref{fig:residualsample}. This set of points is used to define a second network, 
\begin{equation}
(x,t) \longrightarrow \delta \mathcal{P}=(\delta\rho(x^\mu), \delta p(x^\mu), \delta v^i(x^\mu), \delta B^i(x^\mu))
\end{equation}
which learns an Alfv\'en-like perturbation that partially compensates PDE violations.

\begin{figure}[t]
\centering
\includegraphics[width=\linewidth]{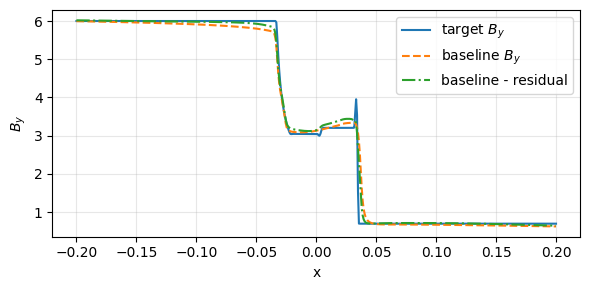}
\caption{Residual output after 5000 training steps, illustrating the residual-guided correction.}
\label{fig:resout5000}
\end{figure}

At each collocation point we treat \(\delta \mathcal P\) as satisfying the linearized inhomogeneous system
\begin{equation}
\mathcal{M}^{a}{}_{b}(\mathcal{P})\,\partial_{t}\,\delta \mathcal{P}^{b}
{+}
\mathcal{A}^{a}{}_{b x}(\mathcal{P})\,\partial_{x}\,\delta \mathcal{P}^{b}
{+}
\mathcal{S}^{a}{}_{b}(\mathcal{P})\,\delta \mathcal{P}^{b}
{=}
\mathcal{R}^{a}(\mathcal{P})\,,
\end{equation}
where \(\mathcal P\) denotes the background solution from the baseline PINN, \(\mathcal M\) is the time Jacobian, \(A_x\) the spatial Jacobian, and $\mathcal{S}(\mathcal{P}) = \partial_t \mathcal{M}(P) + \partial_x \mathcal{A}_x(\mathcal{P})$. The correction network is trained to minimize the mismatch between the two sides of the above equation, ensuring that the corrected surrogate \(\mathcal{P} - \delta \mathcal{P}\) better satisfies the Jacobian-form PDE. This is depicted in Fig.~\ref{fig:resout5000}.

\section{Discussion and Outlook}\label{sec:outlook}

The RMHD-NN framework combines a primitive-variable Jacobian formulation of RMHD with standard PINNs, an aggressive optimizer schedule, and residual-guided correction stages. In the current tests (1D shock tube and 2D cylindrical explosion) the method:
\begin{enumerate}
  \item reproduces RMHD dynamics from limited early-time data,
  \item extrapolates to later times using only PDE information,
  \item systematically reduces residuals via correction networks.
\end{enumerate}

Alternative methods of achieving the last point shall also be explored in the future.

The next step in this direction is the implementation of PINN model for GRMHD equations supported on a curved spacetime, where a rotating black hole is of special interest. In this case initial data with early time simulation is harder to generate, but there exists well known libraries. Furthermore, in real world GRMHD applications, the models serve as a prior in Bayesian methods, and one is only required to achieve the accuracy levels within the measurement expectations.

We find it very promising that the overall time of training does not exceed $\sim$ 90 minutes in a single GPU. This suggests perhaps the network can be escalated to higher accuracy. In doing so the role of an optimizer like MUON is crucial, with alternatives including heavier Gauss-Newton methods \cite{Wang:2025unsing}.

Other natural extensions of our implementation include more than one correction stages, iterating the residual-guided procedure.

A more systematic study of stability, generalization, and comparison to traditional RMHD schemes is left for future work.

\section*{Acknowledgements}

This project was partially funded through the Harvard College Research Program (HCRP). We further thank Harvard Research Computing (HRC) for availability of computing resources. We especially thank Richard Qiu for collaboration and guidance at the beginning stages of this project. We thank Mark Goldstein for further discussions and for suggesting the application of a MUON optimizer. Simulation of initial/early time data was done using the Black Hole Accretion Code (BHAC) which is publicly available in \cite{Porth:2017BHAC}.
This publication is funded in part by the Gordon and Betty Moore Foundation, Grant GBMF12987. This work was supported by the Black Hole Initiative, which is funded by grants from the John Templeton Foundation (Grant \#62286) and the Gordon and Betty Moore Foundation (Grant GBMF-8273) --- although the opinions expressed in this work are those of the authors) and do not necessarily reflect the views of these Foundations.

This write-up is based on our publicly available RMHD-NN code repository, see \url{https://github.com/aguevara22/RMHD-NN}.
\newpage
\appendix
\section{RMHD Equations and Jacobians}\label{app:jacobians}
In this Appendix we list the explicit form of the RMHD equations and its Jacobians. 
The RMHD equations model a plasma in fluid state interacting with an electromagnetic field. 
As such, we fundamentally have three first order constraints for evolution. 
In relativistic notation these are $\partial_\mu \mathcal{T}^{\mu \nu}{=}0$, $\partial_{\mu} F^{\mu\nu}{=}0$ and $\partial_{\mu} (\epsilon^{\mu\nu}{}_{\rho\sigma}F^{\rho\sigma}){=}0$, where $\mathcal T^\mu\nu$ is the stress-energy tensor, $F^{\mu\nu}$ is the Maxwell tensor, and $\epsilon^{\mu\nu\rho\sigma}$ is the Levi-Civita symbol. 
These equations each correspond to stress-energy conservation (of a perfect-fluid and electromagnetic 4-momentum), Gauss' law and Ampere's law, and the no-monopole condition and Faraday's law. 
The second equation is typically approximated with an Ohmic relationship, $J^\mu{=}\sigma_{\mu}{}_\nu F^{\mu\nu}$ where $\sigma^{\mu\nu}$ is the conductivity tensor.
The equations simplify drastically in an `ideal' MHD scenario, namely the condition of infinite conductivity (zero-resistivity).
The ideal-MHD condition is equivalent to $F^{\mu\nu} U_{\nu}{=}0$ where $U_{\nu}$ is a relativistic 4-velocity, representing the absence of the electric field in the plasma frame.

The full RMHD system is obtained by grouping together time and spatial derivative in the above equations. In this appendix we will switch to a more conventional notation to describe \eqref{eq:consd}. We will use bolded notation for three-vectors $\boldsymbol{v}=\{v^i\}$ and dyadics $\boldsymbol{ab}=\{a^ib^j\}$. In this form we have,
\begin{equation}
\frac{\partial}{\partial t}\left(\begin{array}{c}
D \\
\boldsymbol{m} \\
\varepsilon \\
\boldsymbol{B}
\end{array}\right)+\nabla \cdot\left(\begin{array}{c}
D \boldsymbol{v} \\
w_t \gamma^2 \boldsymbol{v} \boldsymbol{v}-\boldsymbol{b} \boldsymbol{b}+\mathrm{I}\, p_t \\
\boldsymbol{m} \\
\boldsymbol{v} \boldsymbol{B}-\boldsymbol{B} \boldsymbol{v}
\end{array}\right)^T=\left(\begin{array}{c}
0 \\
\boldsymbol{0} \\
0 \\
\mathbf{0}
\end{array}\right).
\end{equation}
With these equations, we define the conserved quantities,
\begin{equation}
\begin{array}{lcc}
D= & \gamma \rho \\
\boldsymbol{m}= & w_t \gamma^2 \boldsymbol{v}-b^0 \boldsymbol{b} \\
\varepsilon= & w_t \gamma^2-b^0 b^0-p_t
\end{array}
\end{equation}
involving the following functions,
\begin{equation}
\begin{array}{rlc}
b^0 & =\quad \gamma \boldsymbol{v} \cdot \boldsymbol{B} \\
\boldsymbol{b} & =\quad \boldsymbol{B} / \gamma+\gamma(\boldsymbol{v} \cdot \boldsymbol{B}) \boldsymbol{v}\\
w_t&=\rho h+ b^2  \\
p_t&=p+\frac{b^2}{2}\\
\gamma&=1/\sqrt{1-v^2}\,\,,\,\, v^2 := v_x^2 + v_y^2 + v_z^2.\label{eq:lrfa}
\end{array}
\end{equation}
The currents/fluxes $\boldsymbol{J}^a$ correspond to a $3\times 8$ matrix, with $a=1,\ldots,8$. For instance
\begin{equation}
    \boldsymbol{J}^1=D\boldsymbol{v}\,,\quad \boldsymbol{J}^2=w_t \gamma^2 \boldsymbol{v} \,v_x-\boldsymbol{b}\, b_x+ \hat{\boldsymbol{i}}\,\, p_t \,,\quad \boldsymbol{J^3}=\ldots
\end{equation}
etc. Recall that $\gamma$ is the Lorentz factor of the plasma and that $b^\mu=(b^0,\boldsymbol{b})$ is the magnetic field in such frame. This yields the useful identity for the 4-norm $b^2\equiv b^\mu b_\mu$,
\begin{equation}
    b^2 = \boldsymbol{B}^2 / \gamma^2+(\boldsymbol{v} \cdot \boldsymbol{B})^2
\end{equation}

For completeness we further summarize the Jacobian formulation used throughout this work.
We adopt the primitive variables
\begin{equation}
P = (\rho,\, v_x,\, v_y,\, v_z,\, B_x,\, B_y,\, B_z,\, p).
\end{equation}

The Jacobian matrices used in the code (Eq.~(3)) are given by
\begin{equation}\label{eq:gdsx}
    M^{ab}= \frac{\partial U^a}{\partial P^b}\,,\quad \boldsymbol{A}^{ab}=\frac{\partial{\boldsymbol{J}^a}}{\partial P^b}
\end{equation}
where $a,b=1,\ldots ,8$ and $\boldsymbol{A}=(A^x,A^y,A^z)$. For illustration purposes we quote here the explicit form of the full $8\times 8$ matrix $M$. Recalling the enthalpy of the $\Gamma$-gas
\begin{equation}
\rho h = \rho + p+u \,,\quad p=(\Gamma-1)u
\end{equation}
we find
\begin{center}
\newcommand{\ginv}{\frac{1}{\gamma^2}}
\resizebox{1.0\textwidth}{!}{$
M(P) =
\begin{pmatrix}
\gamma
 & \gamma^3\rho v_x
 & \gamma^3\rho v_y
 & \gamma^3\rho v_z
 & 0 & 0 & 0 & 0
\\[4pt]
\gamma^2 v_x
 & h\gamma^2{+}B^2{-}B_x^2{+}2h\gamma^4 v_x^2
 & -B_xB_y{+}2h\gamma^4 v_x v_y
 & -B_xB_z{+}2h\gamma^4 v_x v_z
 & -B_y v_y{-}B_z v_z
 & 2B_y v_x{-}B_x v_y
 & 2B_z v_x{-}B_x v_z
 & \dfrac{\Gamma\,\gamma^2 v_x}{\Gamma-1}
\\[4pt]
\gamma^2 v_y
 & -B_xB_y{+}2h\gamma^4 v_x v_y
 & h\gamma^2{+}B^2{-}B_y^2{+}2h\gamma^4 v_y^2
 & -B_yB_z{+}2h\gamma^4 v_y v_z
 & -B_y v_x{+}2B_x v_y
 & -B\!\cdot\!v{+}B_y v_y
 & 2B_z v_y{-}B_y v_z
 & \dfrac{\Gamma\,\gamma^2 v_y}{\Gamma-1}
\\[4pt]
\gamma^2 v_z
 & -B_xB_z{+}2h\gamma^4 v_x v_z
 & -B_yB_z{+}2h\gamma^4 v_y v_z
 & h\gamma^2{+}B_x^2{+}B_y^2{+}2h\gamma^4 v_z^2
 & -B_z v_x{+}2B_x v_z
 & -B_z v_y{+}2B_y v_z
 & -B\!\cdot\!v{+}B_z v_z
 & \dfrac{\Gamma\,\gamma^2 v_z}{\Gamma-1}
\\[4pt]
0 & 0 & 0 & 0 & 1 & 0 & 0 & 0
\\
0 & 0 & 0 & 0 & 0 & 1 & 0 & 0
\\
0 & 0 & 0 & 0 & 0 & 0 & 1 & 0
\\[4pt]
\gamma^2
 & -(B\!\cdot\!v)\,B_x{+}(B^2{+}2h\gamma^4)v_x
 & -(B\!\cdot\!v)\,B_y{+}(B^2{+}2h\gamma^4)v_y
 & -(B\!\cdot\!v)\,B_z{+}(B^2{+}2h\gamma^4)v_z
 & B_x\!\left(2{-}\ginv\right){-}(B\!\cdot\!v)v_x
 & B_y\!\left(2{-}\ginv\right){-}(B\!\cdot\!v)v_y
 & B_z\!\left(2{-}\ginv\right){-}(B\!\cdot\!v)v_z
 & -1{+}\dfrac{\Gamma\,\gamma^2}{\Gamma-1}
\end{pmatrix}
$}
\end{center}

The matrices $A^x(P)$ and $A^y(P)$ are also $8\times 8$, we refer the reader to the exact definitions in our publicly available code \texttt{jacobians.py}. For computational efficiency they are computed via symbolic differentiation using the software \texttt{Mathematica}, but can be shown to match, numerically and precisely, with the backpropagating gradient of the \texttt{pytorch} environment (obtained from their definition \eqref{eq:gdsx}). 

All together, this system determines the characteristic speeds of the linearized RMHD system via the eigenvalues of $M^{-1}A^a k^a$, which include the relativistic fast/slow magnetosonic and Alfvén families.

\bibliographystyle{aasjournal}
\bibliography{refs}

@article{Kharazmi:2020vpinn,
  author        = {Kharazmi, Ehsan and Zhang, Zachary and Karniadakis, George Em},
  title         = {Variational Physics-Informed Neural Networks},
  journal       = {arXiv preprint arXiv:2001.04536},
  year          = {2020},
  eprint        = {2001.04536},
  archivePrefix = {arXiv},
  primaryClass  = {cs.LG},
  url           = {https://arxiv.org/abs/2001.04536}
}

@phdthesis{Anton2008PhD,
  author = {Ant{\'o}n, Luis},
  title = {Magnetohidrodin{\'a}mica relativista num{\'e}rica: Aplicaciones en relatividad especial y general},
  school = {Universitat de Val{\`e}ncia},
  year = {2008}
}

@article{JagtapKawaguchi,
  title   = {Adaptive activation functions accelerate convergence in deep and physics-informed neural networks},
  author  = {Jagtap, Ameya D. and Kawaguchi, Kenji and Karniadakis, George Em},
  journal = {Journal of Computational Physics},
  volume  = {404},
  year    = {2020},
  pages   = {109136},
  doi     = {10.1016/j.jcp.2019.109136}
}

@article{AntonEtAl2010,
  author = {Ant{\'o}n, L. and Miralles, J. A. and Mart{\'i}, J. M. and Ib{\'a}{\~n}ez, J. M. and Aloy, M. A. and Mimica, P.},
  title = {Relativistic magnetohydrodynamics: Renormalized eigenvectors and full wave decomposition Riemann solver},
  journal = {Astrophysical Journal Supplement Series},
  volume = {188},
  pages = {1},
  year = {2010},
  eprint = {0912.4692},
  archivePrefix = {arXiv},
  primaryClass = {astro-ph.IM}
}

@article{SchoepeHilditchBugner2018,
  author = {Schoepe, A. and Hilditch, D. and Bugner, M.},
  title = {Revisiting hyperbolicity of relativistic fluids},
  journal = {Physical Review D},
  volume = {97},
  pages = {123009},
  year = {2018},
  eprint = {1712.09837},
  archivePrefix = {arXiv},
  primaryClass = {gr-qc}
}

@article{HilditchSchoepe2019,
  author = {Hilditch, D. and Schoepe, A.},
  title = {Hyperbolicity of divergence cleaning and vector potential formulations of general relativistic magnetohydrodynamics},
  journal = {Physical Review D},
  volume = {99},
  pages = {104034},
  year = {2019}
}

@article{Friedrichs1974,
  author = {Friedrichs, K. O.},
  title = {On the laws of relativistic electromagneto-fluid dynamics},
  journal = {Communications on Pure and Applied Mathematics},
  volume = {27},
  pages = {749},
  year = {1974}
}

@misc{Owkes2024FirstCFDSolver,
  author       = {Owkes, Mark},
  title        = {A guide to writing your first CFD solver},
  year         = {2024},
  month        = apr,
  note         = {PDF lecture note/guide, Montana State University; dated April 11, 2024}
}

@article{FerrerSanchez2024GAPINN,
  title   = {Gradient-annihilated {PINNs} for solving {Riemann} problems: Application to relativistic hydrodynamics},
  author  = {Ferrer-S{\'a}nchez, Antonio and Mart{\'i}n-Guerrero, Jos{\'e} D. and Ruiz de Austri-Baz{\'a}n, Roberto and Torres-Forn{\'e}, Alejandro and Font, Jos{\'e} A.},
  journal = {Computer Methods in Applied Mechanics and Engineering},
  volume  = {424},
  pages   = {116906},
  year    = {2024},
  doi     = {10.1016/j.cma.2024.116906}
}

@book{Toro2009,
  author    = {Toro, Eleuterio F.},
  title     = {Riemann Solvers and Numerical Methods for Fluid Dynamics: A Practical Introduction},
  edition   = {3rd},
  publisher = {Springer},
  year      = {2009},
  address   = {Berlin, Germany}
}

@book{Anile1989,
  author = {Anile, A. M.},
  title = {Relativistic Fluids and Magneto-Fluids: With Applications in Astrophysics and Plasma Physics},
  publisher = {Cambridge University Press},
  address = {New York},
  year = {1989}
}

@article{Dedner:2002glm,
  author  = {Dedner, Andreas and Kemm, Frank and Kr{\"o}ner, Dietmar and Munz, Claus-Dieter and Schnitzer, Thomas and Wesenberg, Matthias},
  title   = {Hyperbolic divergence cleaning for the {MHD} equations},
  journal = {Journal of Computational Physics},
  volume  = {175},
  number  = {2},
  pages   = {645--673},
  year    = {2002},
  doi     = {10.1006/jcph.2001.6961}
}

@article{Gammie:2003HARM,
  author       = "Gammie, Charles F. and McKinney, Jonathan C. and Toth, Gabor",
  title        = "{HARM: A Numerical scheme for general relativistic magnetohydrodynamics}",
  journal      = "Astrophys. J.",
  volume       = "589",
  pages        = "444--457",
  year         = "2003",
  eprint       = "astro-ph/0301509",
  archivePrefix= "arXiv",
  primaryClass = "astro-ph"
}

@article{Sadowski:2013KORAL,
  author       = "Sadowski, Aleksander and Narayan, Ramesh and Tchekhovskoy, Alexander and Zhu, Yucong",
  title        = "{KORAL: A general relativistic radiative magnetohydrodynamics code}",
  journal      = "Mon. Not. Roy. Astron. Soc.",
  volume       = "429",
  pages        = "3533--3550",
  year         = "2013",
  eprint       = "1212.5050",
  archivePrefix= "arXiv",
  primaryClass = "astro-ph.HE"
}

@article{Noble:2006pvs,
  author       = {Noble, Scott C. and Gammie, Charles F. and McKinney, Jonathan C. and Del Zanna, Luca},
  title        = {Primitive variable solvers for conservative general relativistic magnetohydrodynamics},
  journal      = {Astrophys.\ J.},
  year         = {2006},
  volume       = {641},
  pages        = {626--637},
  doi          = {10.1086/500349},
  eprint       = {astro-ph/0512420},
  archivePrefix= {arXiv},
  primaryClass = {astro-ph}
}

@article{Porth:2017BHAC,
  author       = "Porth, Oliver and Olivares, Henok and Mizuno, Yosuke and Younsi, Ziri and Rezzolla, Luciano and Moscibrodzka, Monika and Falcke, Heino and Kramer, Michael",
  title        = "{The Black Hole Accretion Code (BHAC): A 3D GRMHD code for black hole accretion and jet formation}",
  journal      = "Comput. Astrophys. Cosmol.",
  volume       = "4",
  pages        = "1",
  year         = "2017",
  eprint       = "1611.09720",
  archivePrefix= "arXiv",
  primaryClass = "astro-ph.HE"
}

@article{Nagataki:2009grmhd,
  author       = "Nagataki, Shigehiro",
  title        = "{Development of General Relativistic Magnetohydrodynamic Code and Its Application to Central Engine of Long Gamma-Ray Bursts}",
  journal      = "Astrophys. J.",
  volume       = "704",
  pages        = "937–950",
  year         = "2009",
  eprint       = "0902.1908",
  archivePrefix= "arXiv",
  primaryClass = "astro-ph.HE"
}

@article{jordan2024muon,
  author        = {Martens, James and Zhang, Minjia and Chen, Xi and Kaplan, Jared and Li, Xuezhi and Sohl-Dickstein, Jascha},
  title         = {Muon: An optimizer for deep learning},
  journal       = {arXiv preprint arXiv:2405.21055},
  year          = {2024},
  eprint        = {2405.21055},
  archivePrefix = {arXiv},
  primaryClass  = {cs.LG},
  url           = {https://arxiv.org/abs/2405.21055}
}

@article{Dimitropoulos:2025pulsarML,
  author        = "Dimitropoulos, I. and Chaniadakis, E. and Contopoulos, I.",
  title         = "{The 3D pulsar magnetosphere with machine learning: first results}",
  journal       = "Astronomy \& Astrophysics",
  year          = "2025",
  note          = "arXiv:2410.10716v3 [astro-ph.HE]",
  eprint        = "2410.10716",
  archivePrefix = "arXiv",
  primaryClass  = "astro-ph.HE"
}

@article{Wang:2025unsing,
  author       = "Wang, Yongji and Bennani, Mehdi and Martens, James and Racanière, Sébastien and Blackwell, Sam and Matthews, Alex and Nikolov, Stanislav and Cao-Labora, Gonzalo and Park, Daniel S. and Arjovsky, Martin and Worrall, Daniel and Qin, Chongli and Alet, Ferran and Kozlovskii, Borislav and Tomasev, Nenad and Davies, Alex and Kohli, Pushmeet and Buckmaster, Tristan and Georgiev, Bogdan and Gómez-Serrano, Javier and Jiang, Ray and Lai, Ching-Yao",
  title        = "{Discovery of Unstable Singularities}",
  eprint       = "2509.14185",
  archivePrefix= "arXiv",
  primaryClass = "math.AP",
  year         = "2025"
}

@article{Teukolsky:2025charac,
  author       = "Teukolsky, Saul A.",
  title        = "{Characteristic Decomposition for Relativistic Numerical Simulations: I. Hydrodynamics}",
  eprint       = "2511.13836",
  archivePrefix= "arXiv",
  primaryClass = "gr-qc",
  year         = "2025"
}

@ARTICLE{hyerin,
       author = {{Cho}, Hyerin and {Prather}, Ben S. and {Narayan}, Ramesh and {Su}, Kung-Yi and {Natarajan}, Priyamvada},
        title = "{Bridging Scales in Black Hole Accretion and Feedback: Relativistic Jet linking the Horizon to the Host Galaxy}",
      journal = {arXiv e-prints},
         year = 2025,
        month = jul,
          eid = {arXiv:2507.17818},
        pages = {arXiv:2507.17818},
          doi = {10.48550/arXiv.2507.17818},
archivePrefix = {arXiv},
       eprint = {2507.17818},
 primaryClass = {astro-ph.HE},
       adsurl = {https://ui.adsabs.harvard.edu/abs/2025arXiv250717818C}
}

@article{Ricarte_2023,
doi = {10.3847/2041-8213/aceda5},
url = {https://doi.org/10.3847/2041-8213/aceda5},
year = {2023},
month = {aug},
publisher = {The American Astronomical Society},
volume = {954},
number = {1},
pages = {L22},
author = {Ricarte, Angelo and Narayan, Ramesh and Curd, Brandon},
title = {Recipes for Jet Feedback and Spin Evolution of Black Holes with Strongly Magnetized Super-Eddington Accretion Disks},
journal = {The Astrophysical Journal Letters}
}

@ARTICLE{McKinney_Pulsar,
       author = {{McKinney}, Jonathan C.},
        title = "{Relativistic force-free electrodynamic simulations of neutron star magnetospheres}",
      journal = {\mnras},
         year = 2006,
        month = may,
       volume = {368},
       number = {1},
        pages = {L30-L34},
          doi = {10.1111/j.1745-3933.2006.00150.x},
archivePrefix = {arXiv},
       eprint = {astro-ph/0601411},
 primaryClass = {astro-ph},
       adsurl = {https://ui.adsabs.harvard.edu/abs/2006MNRAS.368L..30M}
}

@article{Joana,
	author = {{Kramer, Joana A.} and {MacDonald, Nicholas R.} and {Paraschos, Georgios F.} and {Ricci, Luca}},
	title = {3D hybrid fluid-particle jet simulations and the importance of synchrotron radiative losses},
	DOI= "10.1051/0004-6361/202450978",
	url= "https://doi.org/10.1051/0004-6361/202450978",
	journal = {A\&A},
	year = 2024,
	volume = 691,
	pages = "A14",
}

@article{Ighina_quasar,
doi = {10.3847/2041-8213/aded0a},
url = {https://doi.org/10.3847/2041-8213/aded0a},
year = {2025},
month = {sep},
publisher = {The American Astronomical Society},
volume = {990},
number = {2},
pages = {L56},
author = {Ighina, Luca and Caccianiga, Alessandro and Connor, Thomas and Moretti, Alberto and Pacucci, Fabio and Reynolds, Cormac and Afonso, José and Arsioli, Bruno and Belladitta, Silvia and Broderick, Jess W. and Dallacasa, Daniele and Della Ceca, Roberto and Haardt, Francesco and Lambrides, Erini and Leung, James K. and Lupi, Alessandro and Matute, Israel and Rigamonti, Fabio and Severgnini, Paola and Seymour, Nick and Tavecchio, Fabrizio and Vignali, Cristian},
title = {X-Ray Investigation of Possible Super-Eddington Accretion in a Radio-loud Quasar at z = 6.13},
journal = {The Astrophysical Journal Letters}
}

@BOOK{Toro2009-ty,
  title     = "Riemann solvers and numerical methods for fluid dynamics",
  author    = "Toro, Eleuterio F",
  publisher = "Springer",
  edition   =  3,
  month     =  apr,
  year      =  2009,
  address   = "Berlin, Germany",
  copyright = "https://www.springernature.com/gp/researchers/text-and-data-mining",
  language  = "en"
}

@article{Ripperda_2019,
doi = {10.3847/1538-4365/ab3922},
url = {https://doi.org/10.3847/1538-4365/ab3922},
year = {2019},
month = {sep},
publisher = {The American Astronomical Society},
volume = {244},
number = {1},
pages = {10},
author = {Ripperda, B. and Bacchini, F. and Porth, O. and Most, E. R. and Olivares, H. and Nathanail, A. and Rezzolla, L. and Teunissen, J. and Keppens, R.},
title = {General-relativistic Resistive Magnetohydrodynamics with Robust Primitive-variable Recovery for Accretion Disk Simulations},
journal = {The Astrophysical Journal Supplement Series}
}

@ARTICLE{NASA_PINN,
  title     = "Neural network reconstruction of plasma space-time",
  author    = "Bard, C and Dorelli, J C",
  journal   = "Front. Astron. Space Sci.",
  publisher = "Frontiers Media SA",
  volume    =  8,
  month     =  sep,
  year      =  2021,
  copyright = "https://creativecommons.org/licenses/by/4.0/"
}

@article{chael_force_free,
    author = {Chael, Andrew},
    title = {Hybrid GRMHD and force-free simulations of black hole accretion},
    journal = {Monthly Notices of the Royal Astronomical Society},
    volume = {532},
    number = {3},
    pages = {3198-3221},
    year = {2024},
    month = {07},
    issn = {0035-8711},
    doi = {10.1093/mnras/stae1692},
    url = {https://doi.org/10.1093/mnras/stae1692},
    eprint = {https://academic.oup.com/mnras/article-pdf/532/3/3198/58621450/stae1692.pdf},
}

@article{RAISSI2019686,
title = {Physics-informed neural networks: A deep learning framework for solving forward and inverse problems involving nonlinear partial differential equations},
journal = {Journal of Computational Physics},
volume = {378},
pages = {686-707},
year = {2019},
issn = {0021-9991},
doi = {https://doi.org/10.1016/j.jcp.2018.10.045},
url = {https://www.sciencedirect.com/science/article/pii/S0021999118307125},
author = {M. Raissi and P. Perdikaris and G.E. Karniadakis}
}

@article{self_similar_euler,
  title = {Asymptotic Self-Similar Blow-Up Profile for Three-Dimensional Axisymmetric Euler Equations Using Neural Networks},
  author = {Wang, Y. and Lai, C.-Y. and G\'omez-Serrano, J. and Buckmaster, T.},
  journal = {Phys. Rev. Lett.},
  volume = {130},
  issue = {24},
  pages = {244002},
  numpages = {6},
  year = {2023},
  month = {Jun},
  publisher = {American Physical Society},
  doi = {10.1103/PhysRevLett.130.244002},
  url = {https://link.aps.org/doi/10.1103/PhysRevLett.130.244002}
}

@misc{wang_pinn_guide,
      title={An Expert's Guide to Training Physics-informed Neural Networks}, 
      author={Sifan Wang and Shyam Sankaran and Hanwen Wang and Paris Perdikaris},
      year={2023},
      eprint={2308.08468},
      archivePrefix={arXiv},
      primaryClass={cs.LG},
      url={https://arxiv.org/abs/2308.08468}, 
}

@ARTICLE{LBFGS,
  title     = "On the limited memory {BFGS} method for large scale optimization",
  author    = "Liu, Dong C and Nocedal, Jorge",
  journal   = "Math. Program.",
  publisher = "Springer Science and Business Media LLC",
  volume    =  45,
  number    = "1-3",
  pages     = "503--528",
  month     =  aug,
  year      =  1989,
  language  = "en"
}

@misc{ADAM,
      title={Adam: A Method for Stochastic Optimization}, 
      author={Diederik P. Kingma and Jimmy Ba},
      year={2017},
      eprint={1412.6980},
      archivePrefix={arXiv},
      primaryClass={cs.LG},
      url={https://arxiv.org/abs/1412.6980}, 
}

\end{document}